\documentclass[%
 aip,
 jmp,%
 amsmath,amssymb,
 reprint,%
]{revtex4-1}

\usepackage{graphicx}
\usepackage{dcolumn}
\usepackage{bm}
\usepackage{subfigure}

\begin{document}


\title{Temperature dependence of spin-split peaks in transverse electron focusing   }

\author{Chengyu Yan}
 \email{uceeya3@ucl.ac.uk}
\author{Sanjeev Kumar}%
\author{Michael Pepper}
\affiliation{%
	London Centre for Nanotechnology, 17-19 Gordon Street, London WC1H 0AH, United Kingdom\\
}%
\affiliation{
	Department of Electronic and Electrical Engineering, University College London, Torrington Place, London WC1E 7JE, United Kingdom
}%

\author{Patrick See}
\affiliation{%
	National Physical Laboratory, Hampton Road, Teddington, Middlesex TW11 0LW, United Kingdom\\
}%

\author{Ian Farrer}
\altaffiliation[Currently at ]{Department of Electronic and Electrical Engineering, University of Sheffield, Mappin Street, Sheffield S1 3JD, United Kingdom.}
\author{David Ritchie}
\author{Jonathan Griffiths}
\author{Geraint Jones}
\affiliation{%
	Cavendish Laboratory, J.J. Thomson Avenue, Cambridge CB3 OHE, United Kingdom\\
}%

\date{\today}

\begin{abstract}

We present experimental results of transverse electron focusing measurements performed using n-type GaAs. In the presence of a small transverse magnetic field (B$_\perp$), electrons are focused from the injector to detector leading to focusing peaks periodic in B$_\perp$. We show that the odd-focusing peaks exhibit a split, where each sub-peak represents population of a particular spin branch emanating from the injector. The temperature dependence reveals the peak splitting is well defined at low temperature whereas it smears out at high temperature indicating the exchange-driven spin polarisation in the injector is dominant at low temperatures. 

\end{abstract}

\maketitle

\section*{Background}

The electron transport through a quasi one-dimensional (1D) system realised using the two-dimensional electron gas (2DEG) formed at the interface of GaAs/AlGaAs heterostructure has been extensively studied. A 1D system provides an outstanding platform to envisage not only the non-interacting quantum mechanical system where the conductance quantisation\cite{TPA86, DTR88,WVB88} is in the units of n$\times$$\frac{2e^2}{h}$, where n=1,2,3... are different 1D energy subbsands, but also a venue to explore many-body physics\cite{TNS96,SFP08,HTP09,KTS14,YSM17,YSP17}. Recently, the progress in the physics of many-body 1D system has gained momentum due to prediction and experimental demonstration of rich-phases in low-density 1D system leading to incipient Wigner crystallisation\cite{HTP09,KTS14,JKM09}. Moreover the origin of the 0.7 conductance anomaly in the frame work of many-body 1D system is still debated\cite{WB96,WB98,RDJ02,RDJ05,BCF01}. The 0.7 anomaly has two major features: first, in the presence of in-plane magnetic field, the 0.7 anomaly evolves into 0.5$\times$$\frac{2e^2}{h}$ plateau, which indicates it is spin-related\cite{TNS96}; second, the 0.7 anomaly was found to weaken (strengthen) with decreasing (increasing) temperature\cite{TNS96}. These remarkable observations have led to a volume of theoretical and experimental attempts to probe the intrinsic spin polarisation associated with the 0.7 anomaly, however there is no consensus as such on the origin of this anomaly\cite{WB96,WB98,RDJ02,RDJ05,BCF01}. Therefore, to shed more light on the 0.7 anomaly, it is essential to perform a direct measurement on the spin polarisation within a 1D channel.

\begin{figure*}

\includegraphics[height=2.5in,width=6.4in]{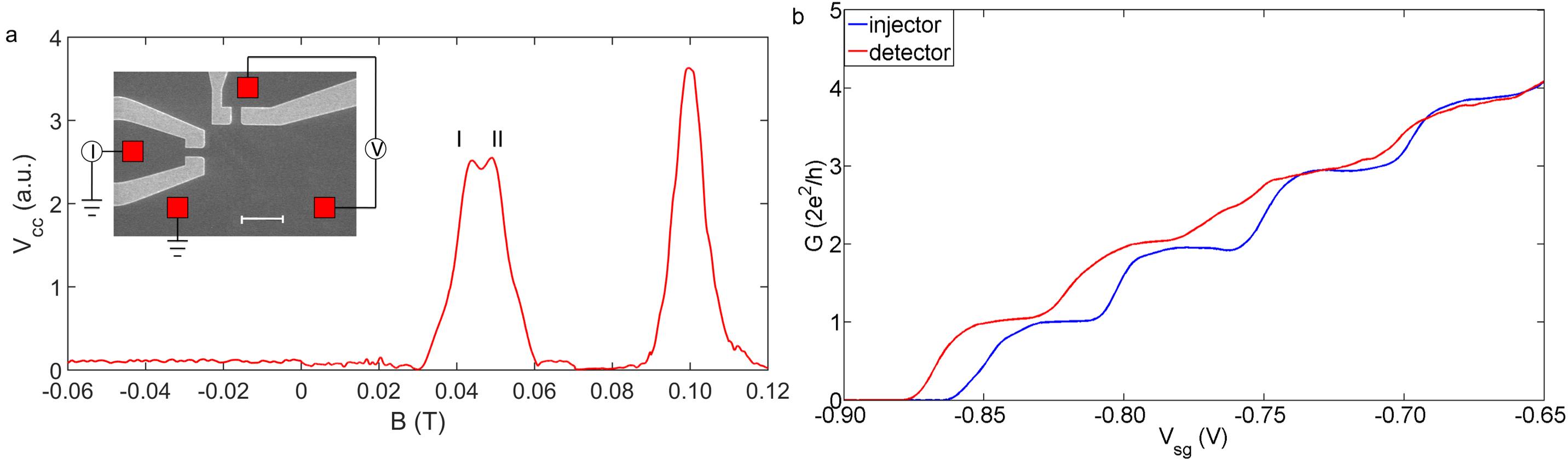}

	\caption{\textbf{The experiment setup and device characteristic.} (a) A representative plot of transverse electron focusing with both the injector and detector set to G$_0$ (2e$^2$/h). V$_{cc}$ is the voltage drop across the detector. Focusing peaks are well defined with positive magnetic field and the signal is negligible with negative magnetic field. The first peak shows pronounced splitting. The two sub-peaks have been highlighted as peak I and peak II. The inset shows an SEM image of the device. The separation between the injector and detector is 1.5 $\mu$m.  Red squares form the Ohmic contacts whereas two pairs of grey-coloured gates, left and top, form the injector and detector, respectively. The scale bar is 2 $\mu$m. (b) Conductance characteristics of the injector and detector. 
	}           
	\label{fig:BasicInf}
\end{figure*} 

A scheme based on transverse electron focusing (TEF) was proposed to address the spin polarisation\cite{GC04,AGC07}, and was validated in p-type GaAs\cite{LPR06,SC11} and n-type InSb\cite{ARD06}. Within this scheme, the spin polarisation can be extracted from the asymmetry of the two sub-peaks of the first focusing peak. Recently, we showed that injection of 1D electrons whose spins have been spatially separated, can be detected in the form of a split in the first focusing peak, where the two sub-peaks represent the population of detected spin states\cite{YSM16}. In the present work, we report the temperature dependence of spin-split first focusing peak, and analyse the results based on the spin-gap present between the two spin species. 

\section*{Method}
The devices studied in the present work were fabricated from the high mobility two dimensional electron gas (2DEG) formed at the interface of GaAs/Al$_{0.33}$Ga$_{0.67}$As heterostructure. At 1.5 K, the measured electron density (mobility) was 1.80$\times$10$^{11}$cm$^{-2}$ (2.17$\times$10$^6$cm$^2$V$^{-1}$s$^{-1}$), therefore the mean free path is over 10 $\mu$m which is much larger than the electron propagation length. The experiments were performed in a cryofree dilution refrigerator with a lattice temperature of 20 mK  using the standard lockin technique.  The range of temperature dependence measurement was from 20 mK to 1.8 K. 

\begin{figure*}

\includegraphics[height=5.0in,width=6.4in]{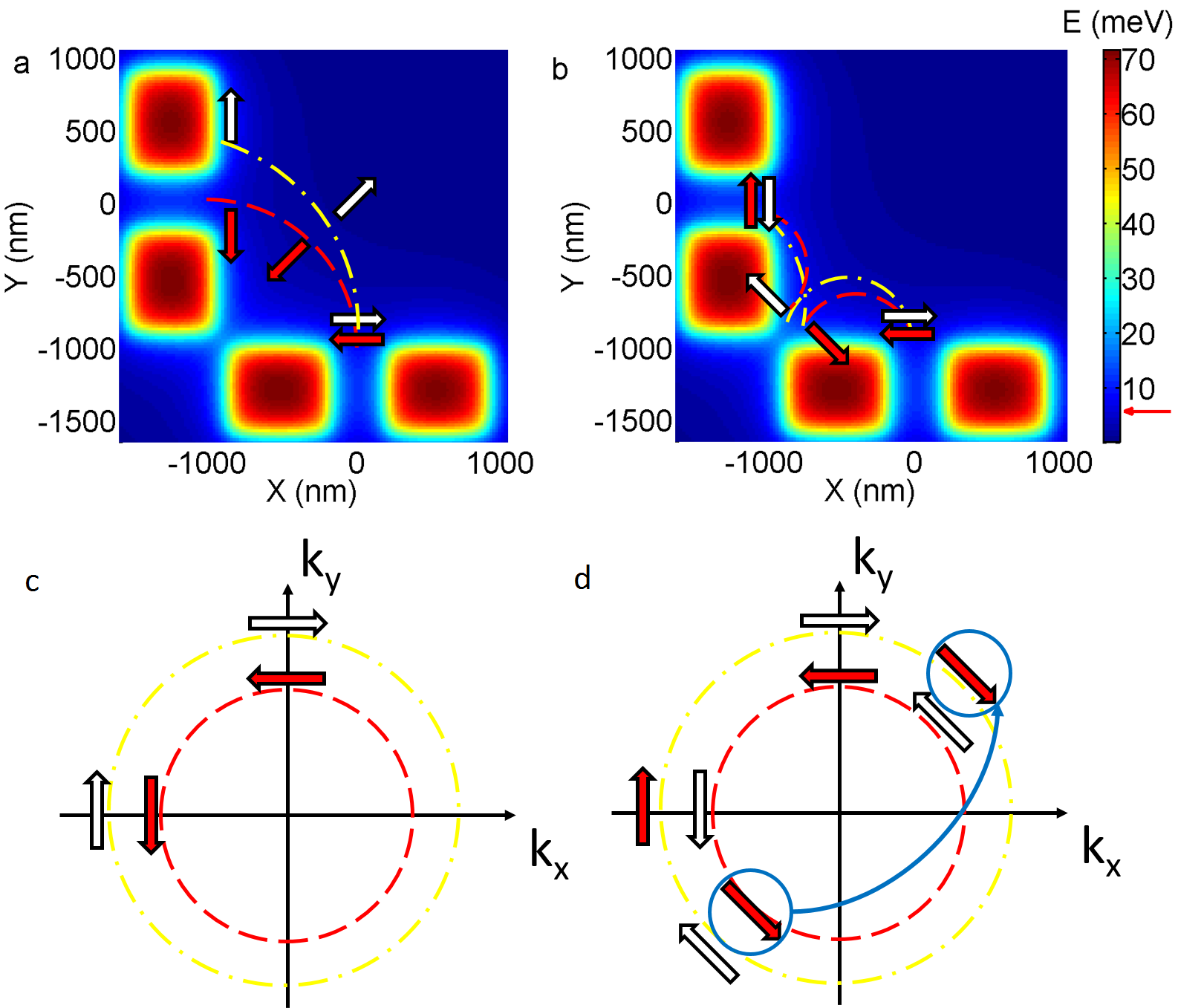}

	\caption{\textbf{Mechanism of peak splitting.}       (a)-(b) Peak splitting in the coordinate-space for first and second focusing peaks, respectively. The red and white arrows represent spin-up and spin-down electrons, the coloured blocks  stand for the electrostatic potential and the red-dashed trace is with smaller cyclotron radius while the yellow-dotted one is with larger cyclotron radius. (c)-(d), Peak splitting in the k-space for first and second focusing peaks, respectively. The electrons travel from (0, k$_y$) to (-k$_x$, 0) anticlockwise in plot (c). In plot (d), the thick blue arrow highlights the transition after reflection at the boundary of electrostatic potential formed between the injector and detector. 
	}           
	\label{fig:T_dep}
\end{figure*}

\section*{Results and discussion}

Figure~1(a) shows the experimental setup along with a typical focusing spectrum obtained using the device shown in the inset. The focusing device is specially designed so that the injector and detector can be controlled separately to avoid a possible cross-talking between them\cite{YSM16, RMP02,HVH89}. The quantum wire used for the injector and detector has a width (confinement direction) of 500 nm and length (current flow direction) of 800 nm. Both the injector and detector show well defined conductance plateaus as shown in Fig.~1(b).  Further details on the device are given in the caption of Fig. 1. 

With negative magnetic field, the measured signal is almost zero because electrons bend into the opposite direction and thus miss the detector. It is also evident that the Shubnikov-de Haas oscillation and quantum Hall effect do not contribute to the observation. In the presence of a small positive transverse magnetic field B$_{\perp}$ electrons are focused from the injector to detector leading to focusing peaks periodic in B$_{\perp}$ while the detected signal is negligible at the negative magnetic field end. The calculated periodicity of 60 mT using the relation\cite{HVH89}, $$B_{focus}=\frac{\sqrt{2}\hbar k_F}{eL} \eqno(1)$$ is in good agreement with the experimental result. Here \textit{e} is the elementary charge and \textit{$\hbar$} is the reduced Planck constant, \textit{L} is the separation between the injector and detector (in the 90$^{\circ}$ focusing device geometry, this is the separation along the diagonal direction).  In addition to the periodic focussing peak which is a manifestation of the semi-classical electron cyclotron orbit, it is interesting to notice the splitting of odd-numbered focusing peaks. It is suggested that this anomalous splitting of odd-numbered focusing peaks arises from the spin-orbit interaction (SOI)\cite{GC04,AGC07} and has been successfully observed in GaAs hole gas\cite{LPR06,SC11} and InSb electron gas\cite{ARD06}. We recently demonstrated splitting of odd-numbered focusing peaks in n-GaAs\cite{YSM16} where a longer quantum wire possessing partially polarised and spatially separated 1D electrons was used to inject the polarised 1D electrons  into the 2D regime and subsequently measured across the detector in the form of a split in the first focusing peak. Here we are interested in investigating the thermal effect on the spin states within the 1D channel via the transverse electron focusing. We note that the splitting smears out when the thermal energy k$_B$T exceeds 2$\Delta$E ($\Delta$E is the energy difference between the two spin branches) agreeing with the theoretical prediction\cite{AGC07}. 

\begin{figure*}

  	\includegraphics[height=4.0in,width=6.4in]{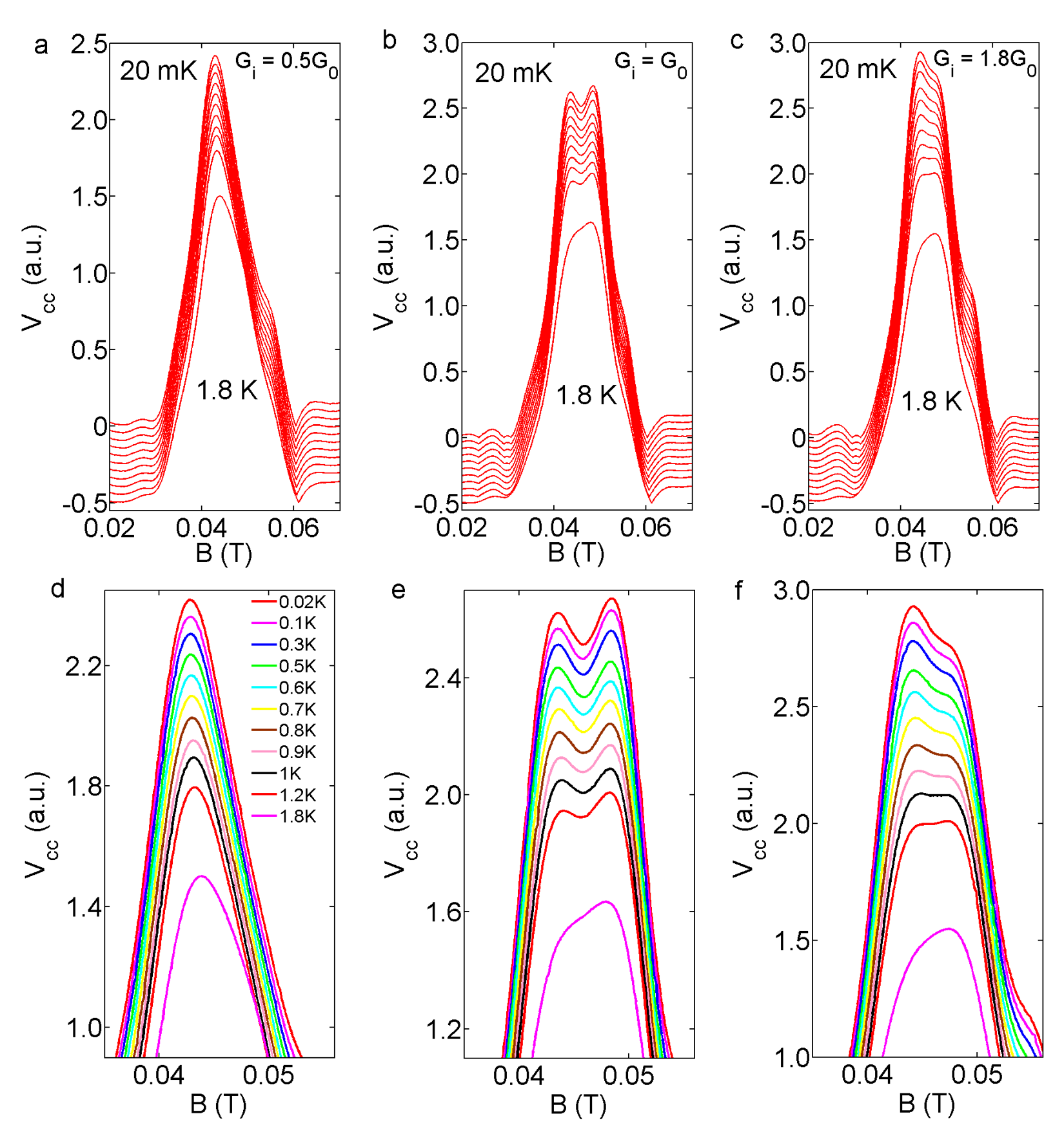}

	\caption{\textbf{Temperature dependence of TEF.} (a)-(c) The injector was set to 0.5$G_0$, G$_0$ and 1.8G$_0$, respectively. The lattice temperature was incremented from 20 mK (top trace) to 1.8 K (bottom trace).  Data have been offset vertically for clarity. (d)-(f), zoom-in of the data in (a)-(c).
	}           
	\label{fig:fitting}
\end{figure*} 

Before we discuss the temperature dependence effect, it is important to understand the mechanism responsible for the observed peak splitting. Figure 2(a) and (b) show the potential profile of the split gates forming the injector (bottom pair) and the detector (left pair). In the presence of SOI, the two spin species follow different cyclotron radii as shown in Fig.~2(a) thus resulting in two sub-peaks in the first focusing peak. However, the situation is different for the second focusing peak where a scattering at the boundary of electrostatic potential  created by the split gates is involved as shown in Fig.~2(b). In this case, a spin-up electron (red arrow in the colourplots) initially follows a smaller cyclotron radius while it possesses a larger radius after the scattering\cite{GC04,AGC07} and vice-versa for the spin-down electron (white arrow), thus the two spin species re-join at the detector. The underneath reasoning for the peak splitting can be found in the k-space in Fig.~2(c) and (d). Here we assume the spin-orbit interaction is of Rashba-type, however, the analysis holds valid for Dresselhaus effect in bulk as well. For the first focusing peak (Fig.~2(c)), the two spin-species travel from (0, k$_y$) to (-k$_x$, 0) along different Fermi surfaces. For the second focusing peak (Fig.~2(d)), the same argument holds true before the scattering, however, the momentum changes its sign while the spin orientation remains preserved after the scattering\cite{GC04}. Therefore, a spin-up electron (red arrows) initially occupying the inner Fermi surface hops to the outer Fermi surface after the scattering to guarantee both the sign of the momentum and the spin orientation are in the correct order (the hopping is highlighted by the thick blue arrow in Fig.~2(d)) and vice-versa for the spin-down electron. The cyclotron radius is proportional to the momentum, so that the alternation in cyclotron radius occurs in the coordinate-space as a consequence of hopping between two Fermi surfaces which leads to a single second focusing peak. 

\begin{figure*}

  	\includegraphics[height=4.0in,width=6.4in]{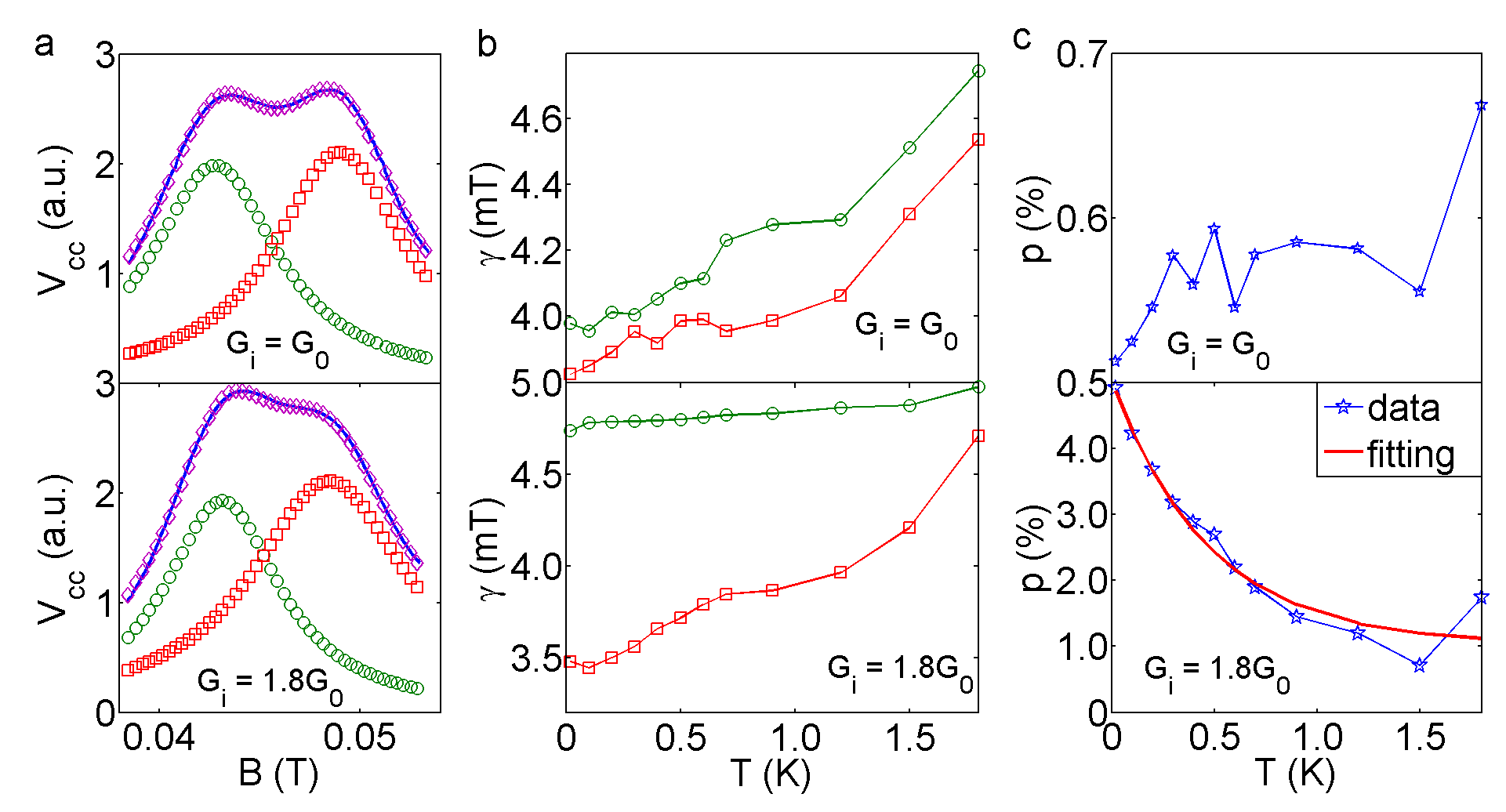}

	\caption{\textbf{Analysis of the temperature dependence data.}  (a) Reconstructing the first focusing peak with two Lorentzian peaks at 20 mK. The solid blue line is the experimental data, the green-round marker is the fit for peak I and red-square marker is the fit for peak II and the magenta-diamond marker highlights the reconstructed focusing peak. (b) FHWM, $\gamma$ as a function of temperature; the sub-peaks broaden with increasing temperature in both cases. The markers represent the same meaning as in plot (a). (c) The polarisation measured with G$_i$ = G$_0$ fluctuates around 0.6$\%$. On the other hand, the polarisation measured with G$_i$ = 1.8G$_0$ follows an exponential decay.
	}           
	\label{fig:fitting}
\end{figure*} 

Figure~3(a)-(c) show the temperature dependence of focusing results with injector set to 0.5G$_0$, G$_0$ and 1.8G$_0$, respectively, where the lattice temperature is incremented from 20 mK (the electron temperature is calibrated to be around 70 mK) to 1.8 K,  and Fig.~3(d)-(f) shows the zoom-in of the data in Fig.~3(a)-(c), respectively. For G$_i$ = 0.5G$_0$ (Fig.~3(a)) a single peak is observed (as only one spin-subband is occupied), which broadens gradually at higher temperature. In addition, the focusing peak shifts towards the center of the spectrum and becomes more symmetric at higher temperature (see the bottom trace, T = 1.8 K, Fig.~3(a) and (d)). This may be due to a possible electron transition between the two spin-subbands at relatively high temperature.  In comparison, for G$_i$ = G$_0$ (Fig.~3(b)) the sub-peaks, each representing a spin-state, are present from 20 mK up to 1.2 K. However, the dip in the first focusing peak leading to two sub-peaks smears out at 1.8 K (Fig.~3(b) and (e)). With G$_i$ set to 1.8G$_0$  (Fig.~3(c)), the splitting is not well resolved and the left sub-peak (I) dominates the spectrum. We note that on increasing the temperature the peak I gradually reduced in amplitude to result in an asymmetric first focusing peak at 1.8 K. In n-type InSb, the splitting was pronounced even at 10 K, which is consistent with the fact the peak splitting was around 60 mT, an indication of strong SOI in InSb\cite{ARD06}, which is one order larger than the peak splitting of 5.5 mT measured in the present case. 

To extract the peak width and amplitude accurately considering the two sub-peaks may partially overlap with each other, we use two Lorentzian peaks to reconstruct the experimental data as shown in Fig.~4(a) using the relation, $$ A(B) = \sum_{i=1,2} A_i \times \frac{\gamma_i^2}{\gamma_i^2+(B-B_i)^2} \eqno(2)$$ where A$_i$ is the amplitude of the peak $i$ ($i$ =1, 2 for peak I and peak II, respectively), $\gamma_i$ denotes the full width at half maximum (FWHM), $B_i$ is the center of the peak. Two noticeable results can be extracted from the fitting: First, it is seen from Fig.~4(b) that $\gamma$ (see caption of Fig.~4 for details on traces  and symbols representing peak I and peak II) for both peak I and peak II increases with rising temperature regardless of the injector conductance which indicates the thermal broadening of the sub-peaks prevents the observation of peak splitting at high temperature. It may be noted that peak I for G$_i$ = 1.8G$_0$ is relatively robust against temperature compared to other peak (both peaks of G$_0$ and peak II of 1.8G$_0$). Second, the measured spin polarization $p$ ($p= |\frac{A_1-A_2}{A1+A_2}|$ ) with G$_i$ = G$_0$ fluctuates around 0.6$\%$ and shows no explicit temperature dependence which agrees with the fact that spin polarisation at conductance plateau should remain at 0 regardless of temperature (Fig.~4(c), upper plot). On the other hand, when G$_i$ is set to 1.8G$_0$, the extracted spin polarisation decays from 5$\%$ to 0.8$ \%$ (Fig.~4(d), lower plot) following the relation\cite{BCF01}, $$ p = \alpha exp(-\frac{k_B T}{\Delta E}) + c \eqno(3)$$ where $\alpha$ is a prefactor accounting for the amplitude, $k_B$ is the Boltzmann constant, $\Delta E$ is the energy difference between the two spin-branches and \textit{c} accounts for the small residual value arises from the uncertainty in the experiment. We extracted the value of $\Delta E$ to be around 0.041 meV (corresponding to 0.5 K). The theory\cite{AGC07} predicts the splitting should persist until $k_BT$ exceeds 2$\Delta E$ (i.e. 1 K in our case) which agrees resonablely well with our result that the peak splitting is observable up to 1.2 K.

\section*{Conclusion}
In conclusion, we showed the temperature dependence of the transverse electron focusing where the contribution of the two spin states manifested as two sub-peaks in the first focusing peak. It was observed that the peak splitting is well defined from 20 mK up to 1.2 K and beyond this temperature the peak splitting smeared out. Moreover, the focusing peak has a tendency to become more symmetric  at higher temperature indicating a possible equilibrium between the two spin branches due to thermal excitation.

The work is funded by the Engineering and Physical Sciences Research Council (EPSRC), UK.

\end{document}